\documentclass[twocolumn,showpacs,superscriptaddress,prl]{revtex4}

\usepackage[latin9]{inputenc}
\usepackage{color}
\usepackage{verbatim}
\usepackage{amsmath}
\usepackage{amssymb}
\usepackage{graphicx}
\usepackage{esint}
\usepackage{hyperref}

\makeatletter
\@ifundefined{textcolor}{}
{%
 \definecolor{BLACK}{gray}{0}
 \definecolor{WHITE}{gray}{1}
 \definecolor{RED}{rgb}{1,0,0}
 \definecolor{GREEN}{rgb}{0,1,0}
 \definecolor{BLUE}{rgb}{0,0,1}
 \definecolor{CYAN}{cmyk}{1,0,0,0}
 \definecolor{MAGENTA}{cmyk}{0,1,0,0}
 \definecolor{YELLOW}{cmyk}{0,0,1,0}
}

\usepackage{mathrsfs}

\draft

\begin{document}
\title{Quantum Dynamical Phase Transition in a Spin-Orbit Coupled Bose Condensate}
\author{Jeffrey Ting Fung Poon}
\affiliation{International Center for Quantum Materials, School of Physics, Peking University, Beijing 100871, China}
\affiliation{Collaborative Innovation Center of Quantum Matter, Beijing 100871, China}
\author{Xiong-Jun Liu
\footnote{Corresponding author: xiongjunliu@pku.edu.cn}}
\affiliation{International Center for Quantum Materials, School of Physics, Peking University, Beijing 100871, China}
\affiliation{Collaborative Innovation Center of Quantum Matter, Beijing 100871, China}

\begin{abstract}
Spin-orbit coupled bosons can exhibit rich equilibrium phases at low temperature and in the presence of particle-particle interactions. In the case with a 1D synthetic spin-orbit interaction, it has been observed that the ground state of a Bose gas can be a normal phase, stripe phase, or magnetized phase in different experimentally controllable parameter regimes. The magnetized states are doubly degenerate and consist of a many-particle two-state system. In this work, we investigate the nonequilibrium quantum dynamics by switching on an external perturbation to induce resonant couplings between the magnetized phases, and predict the novel quantum spin dynamics which cannot be obtained in the single-particle systems. In particular, due to particle-particle interactions, the transition of the Bose condensate from one magnetized phase to the other is forbidden when the strength of external perturbation is less than a critical value, and a full transition can occur only when the perturbation exceeds such critical strength. This phenomenon manifests itself a quantum dynamical phase transition, with the critical point behavior being exactly solvable. From the numerical simulations and exact analytic studies we show that the predicted many-body effects can be well observed with the current experiments.
\end{abstract}
\date{\today }
\maketitle

\indent

\section*{Introduction}
In quantum mechanics, a resonant quantum oscillation can occur between two single-particle states when the system is subject to external perturbation which drives resonant couplings between the states~\cite{Feynman}. The full transition from one quantum state to another can occur after one-half period evolution, no matter how weak the resonant coupling is. Such two-state mechanism has been at the heart of understanding fundamental quantum dynamics and developing applications in atomic, molecular, optical, and condensed matter systems. A Bose-Einstein condensate (BEC) of many atoms can be described with a global wave function, which resembles a single-particle quantum state~\cite{BECbook}. In the absence of particle-particle interactions, the quantum dynamics of a BEC are similar as those in a single-particle system.

Recently, a breakthrough progress for cold atoms is the experimental realization of synthetic spin-orbit (SO) coupling for (pseudo)spin-1/2 bosons~\cite{Ian2011}
and fermions~\cite{Zhang2012,Martin2012}, which has
generated exciting pursuit of exotic physics including the novel SO effects~\cite{Lewenstein2005,Fleischhauer2005,Liu2007,Zhu2006,Liu2009,Shenoy2011,Naturecom2011,Wu2012,Ohberg2013,Wei2013,Zhou2013,Sherman2014,Galitski2013} and topological phases~\cite{Chuanwei2008,Sato2009,LiuPRB2009,Zhu2011,Liu2013,Liu2014,Goldman2014,Hasan2010,Zhang2011}, and open new possibilities to probe novel quantum spin dynamics. So far the realized SO interaction is of the one-dimensional (1D) form, and is generated through two-photon Raman process which couples the two spin states and transfers momentum between them simultaneously~\cite{Fleischhauer2005,Liu2009}. The SO coupling brings a rich structure in the single particle spectra of which the lowest energy subband can be in the single-well or double-well form by tuning the Raman coupling strength~\cite{Liu2009}. For bosons, the double-well dispersion relation is particularly interesting, for it has two degenerate minima with opposite finite momenta in the single-particle spectrum. As a consequence of the competition between the kinetic and interaction energies, at low temperature the bosons may condense in a fixed finite momentum state (magnetized phase), which is of two-fold degeneracy, or in the superposition of two degenerate states with opposite finite momenta (stripe phase)~\cite{Zhai2010,Wu2011,Ho2011,Li2012}. These novel phases have been recently observed in the experiment with $^{87}$Rb atoms~\cite{Chen2014}. Rather than being single-particle states, the two magnetized phases with opposite finite momenta consist of a degenerate many-particle two-state system. This brings about an interesting question that what nontrivial dynamics can be expected by applying an external perturbation to couple resonantly such two many-particle ground states?

{\bf Results.}--In this work, we investigate the transitions between different many-particle states of a SO coupled BEC induced by external perturbation, and uncover novel quantum spin dynamics which cannot be obtained from single-particle systems. The main results are outlined below. Starting from the magnetized phase with the momentum $k_x=k_m$, an external perturbation is switched on to drive transition from the initial phase to the one at $k=-k_m$. (i) A critical value $V_{p,\rm crit}$, which depends on the particle-particle interactions, is predicted for the perturbation. When the strength of perturbation is below the critical value, the maximum transition ratio, given by $|\beta^2_{\rm max}|$, is always less than $1/2$. (ii) By increasing the perturbation strength to be above $V_{p,\rm crit}$, we find that the maximum transition ratio sudden jumps from $|\beta^2_{\rm max}|<1/2$ to $|\beta^2_{\rm max}|=1$, rendering a novel quantum dynamical phase transition~\cite{dynamicaltransition1,dynamicaltransition2,dynamicaltransition3}. (iii) The time of transition between the two ground phases diverges logarithmically as the perturbation approaches the critical point, while in the case with the perturbation far above the critical magnitude, the transition becomes similar to the quantum dynamics of a single-particle two-state system. These predictions can be observed in the experiment by measuring the time evolution of spin polarization of the condensate.

\section*{Magnetized Phases and Induced Couplings}

The system we consider is a spin-1/2 BEC of $^{87}$Rb atoms with 1D synthetic SO interaction generated by two-photon Raman coupling~\cite{Ian2011,Chen2012}. In the presence of the particle-particle interactions, the Hamiltonian of the system is given by (let $\hbar=m=1$ for simplicity)
\begin{eqnarray}
H_{\rm BE} &=& H_0+H_{\rm int}, \nonumber
\end{eqnarray}
where the single-particle term $H_0$ includes 1D SO coupling along $x$ direction and $H_{\rm int}$ is the interacting term
\begin{eqnarray}
H_0&=&\sum_{ss'=\uparrow,\downarrow}\int d^3 \bold r\psi_s^\dag
\bigr(-\frac{\bold \nabla^2_{\bold r}}{2}+ik_0\partial_x\sigma_z+\frac{\Omega}{2}\sigma_x\bigr)_{ss'}
\psi_{s'},\nonumber\\
H_{\rm int}&=&\int d^3 \bold r\frac{g_s}{2} \left( \psi^\dagger_\downarrow \psi^\dagger_\downarrow \psi_\downarrow \psi_\downarrow + \psi^\dagger_\uparrow \psi^\dagger_\uparrow \psi_\uparrow \psi_\uparrow \right)\\
&&+\int d^3 \bold rg_a \psi^\dagger_\downarrow \psi^\dagger_\uparrow \psi_\uparrow \psi_\downarrow.\nonumber
\label{eqn:Hamiltonian}
\end{eqnarray}
Here the field operator $\psi_{s}(\bold r)$ ($\psi_s^\dag(\bold r)$) annihilates (creates) a particle at position $\bold r$ and in the spin state $s=\uparrow,\downarrow$, the two-photon Raman coupling strength $\Omega$ serves as an effective Zeeman field along $x$ axis, and the SO coefficient $k_0$ is determined by the wave vector of Raman lasers. The interacting coefficients $g_s=(g_{\uparrow\uparrow}+g_{\downarrow\downarrow})/2$ and $g_a=g_{\uparrow\downarrow}$, where $g_{ss'}$ represents the interacting coefficient between atomic species $s$ and $s'$. We have neglected the difference between $g_{\uparrow\uparrow}$ and $g_{\downarrow\downarrow}$, since it can be absorbed as an effective two-photon detuning in the Raman coupling and be compensated by properly tuning the frequencies of Raman lasers in the experiment~\cite{Zhai2015}. Diagonalizing the single-particle Hamiltonian  $H_0$ yields two subbands with energies $E^{\pm}_{\rm kin}=\frac{\bold k^2}{2}\pm\sqrt{k_0^2k_x^2+\frac{\Omega^2}{4}}$. The lower subband energy $E^{-}_{\rm kin}$ has two degenerate minima at $\pm k_{\rm min}=\pm k_0(1-\Omega^2/4k_0^4)^{1/2}$ when $|\Omega|<2k_0^2$, while it has a single minimum at $k_{\rm min}=0$ if $|\Omega|>2k_0^2$~\cite{Liu2009}. The former band structure leads to nontrivial phases for bosons even at zero temperature.

\begin{figure}[h]
\includegraphics[width=1.0\columnwidth]{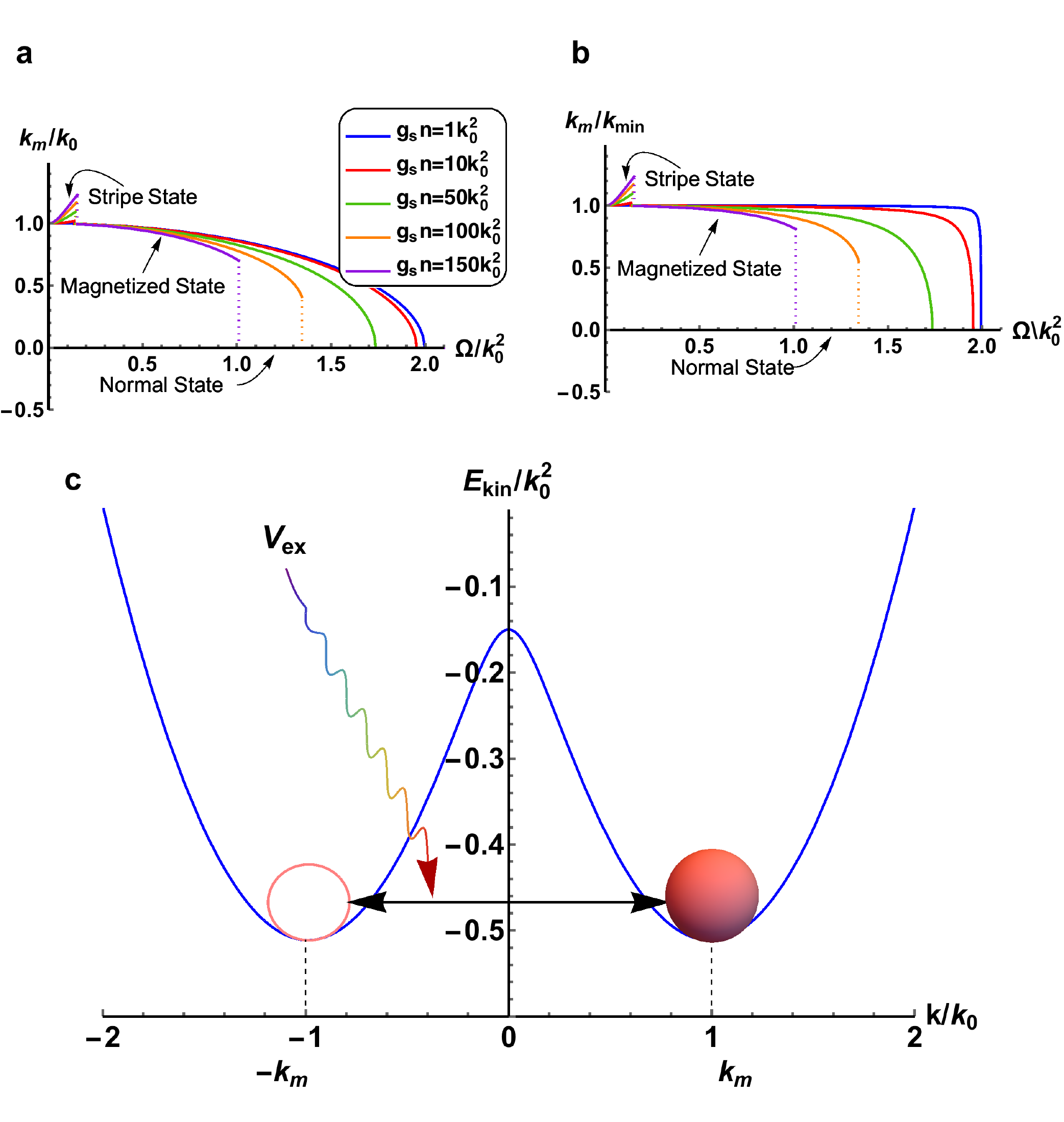}
\caption{{\bf The condensate momentum and induced coupling between magnetized phases.} {\bf a.} The ratio between the condensate momentum $k_m$ and $k_0$ as a function of Raman coupling strength $\Omega$ with different interaction energies $g_sn$ (in the unit of $k_0^2$). {\bf b.} The ratio between the condensate momentum $k_m$ and $k_{\rm min}$. As the Raman coupling $\Omega$ increases, the ratios $k_m/k_0$ and $k_m/k_{\rm min}$ increase for the stripe phase regime, decrease for the magnetized phase regime, and is zero for the normal regime, respectively. {\bf c.} Starting with a magnetized phase, an additional external potential switched on at $t=0$ can drive resonant couplings between the two degenerate magnetized phases.}
\label{phasediagram}
\end{figure}

The Hamiltonian $H_{\rm BE}$ has three distinct ground phases due to the competition between the kinetic and interaction energies which can be controlled in the experiment by manipulating $\Omega$ and the condensate density $n=n_\uparrow+n_\downarrow$, where $n_s$ represents the atomic density of the spin component $s$. In general, the condensate wave function of these ground phases take the following form
\begin{eqnarray}\label{eqn:BECwavefunction1}
\Phi_{\rm BE}(\bold r)=\sqrt{n}\bigr[\alpha u_R(\theta)e^{ik_mx}+\beta u_L(\theta)e^{-ik_mx}\bigr],
\end{eqnarray}
where $u_R=(\cos\theta,-\sin\theta)^T$ and $u_L=(\sin\theta,-\cos\theta)^T$ with $\tan2\theta=\Omega/(2k_0k_m)$ are two single-particle eigenstates of the lower subband of $H_0$ with momenta $k_x=\pm k_m$ (generically different from $\pm k_{\rm min}$) [see Fig.~\ref{phasediagram} (a,b)], $\alpha,\beta$ and $k_m$ are variational parameters and should be determined by minimizing the total energy of the system. The three different phases are separated by two critical Raman strengths $\Omega_{c1,c2}$, which depend on condensate density and satisfy $\Omega_{c1}>\Omega_{c2}$~\cite{Zhai2015}. In the large Raman coupling regime, namely, when $\Omega>\Omega_{c1}$, the total energy minimizes at $k_m=0$, giving the trivial condensate phase. In the small Raman coupling regime with $\Omega<\Omega_{c2}$, the ground phase is a superposition of the states with finite momenta $k_x=\pm k_m$ and $|\alpha|=|\beta|=1/\sqrt{2}$, resulting in the stripe phase. Finally, for the intermediate regime with $\Omega_{c2}<\Omega<\Omega_{c1}$, the system has two degenerate phases at finite momenta $k_x=k_m$ and $k_x=-k_m$, with $\alpha=1,\beta=0$ and $\alpha=0,\beta=1$, respectively, called the magnetized states. These novel phases have been recently observed in the experiment with $^{87}$Rb at temperature of a few tens of nK~\cite{Chen2014}.

The magnetized phases break both the $U(1)$ and $Z_2$ mirror symmetries of the Hamiltonian $H_{\rm BE}$, with the mirror symmetry defined by ${\cal M}_x=i\sigma_x$ and connecting the two magnetized phases at $k_x=\pm k_m$. The main purpose of this work is to study the quantum spin dynamics for this two state system by introducing an external perturbation to couple resonantly the two magnetized phases [see Fig.~\ref{phasediagram} (c)]. Note that the magnetized phases distinct from each other by a finite momentum $\delta k=2k_m$. To induce quantum transitions between such two phases, we switch on a perturbation at the time $t=0$ in the following form
\begin{eqnarray}
V_{\rm ex}(\bold r, t)= \left\{ \begin{array}{ll}
         0, \ \ \ \ \ \ \ \ \ \ \ \ \ \ \ \ \ \ t<0\\
        V_0 \int d^3 \bold r \cos^2 k_mx \left( \psi_\uparrow^\dagger \psi_\uparrow + \psi_\downarrow^\dagger \psi_\downarrow \right), \ t>0.
        \end{array} \right.
\end{eqnarray}
where $V_0$ represents the perturbation strength. The above perturbation is simply an additional lattice potential which can be readily generated by a standing wave light with wave vector $k_R=k_m$. We note that in general $V_{\rm ex}$ can induce couplings between any two states with momentum difference $\pm2k_m$. Thus it may drive transitions from the magnetized phases to other states with e.g. the momenta $k_x=\pm3k_m$. These states, however, have large energy mismatch compared with the energy of the ground phases. Taking into account such transitions does not affect the main results of this work, as we shall discuss later. Below for convenience we first neglect these additional couplings. On the other hand, from Fig.~\ref{phasediagram}(a,b) we read that for magnetized phases $k_m\simeq k_{\rm min}\simeq k_0$, which can facilitate the parameter choice in the real experiment by setting $k_m=k_{\rm min}$ or $k_R=k_0$. The small mismatch of the wave vector only has a tiny quantitative correction to the main results, and will be examined later.

Assume that the system is initialized in the magnetized phase with $k_x=k_m$. The perturbation $V_{\rm ex}$ drives the system to evolve from one magnetized phase to another. Keeping only the transitions between the two ground states, we can describe the time-dependent condensate wave function as $|\Psi_{\rm BE}(\bold r,t)\rangle = [\alpha^*(t) \psi_R^\dag + \beta^*(t) \psi_L^\dag]^N|vac\rangle$, where $N$ is the number of atoms of the BEC, $|vac\rangle$ represents the vacuum state, and the field operator $\psi_R =(\cos\theta\psi_\uparrow-\sin\theta\psi_\downarrow)e^{ikx}
$
and
$\psi_L =(\sin\theta\psi_\uparrow-\cos\theta\psi_\downarrow)e^{-ikx}$, with the initial condition $\alpha(0)=1$ and $\beta(0)=0$. Denoting by $\psi(t)=\alpha(t) \psi_R + \beta(t) \psi_L$, the dynamics of the condensate can be derived according to the following equation:
\begin{eqnarray}\label{eqn:commutator1}
i\frac{d\psi(t)}{dt}=[\psi(t),H_0+H_{\rm int}+V_{\rm ex}].
\end{eqnarray}
The right hand side of the above formula contains three commutators, corresponding to the kinetic energy term, the interacting energy term, and the coupling induced by external perturbation between two magnetized phases, respectively. It will be shown that due to the particle-particle interactions the quantum spin dynamics for the many-particle ground states are fundamentally different from those in the single-particle systems.

\section*{Quantum Oscillation of Magnetized Phases}

\noindent{\bf Equation of Motion.}--We first derive explicitly the equation of motion for the condensate by keeping only the transitions between the two magnetized states. This can be performed by calculating the commutators in Eq.~\eqref{eqn:commutator1} and projecting the results onto the ground states generated by $\psi_R$ and $\psi_L$. The resulted equation in the mean-field approximation governs the quantum spin dynamics for the magnetized phases. The relevant commutators can be verified by (details can be found in the Supplementary Material~\cite{SI})
\begin{equation}
\begin{split}
\left[\psi_{R/L},H_0\right]&=E^-_{\rm kin}\psi_{R/L},\\
\left[\psi_{R/L},V_{\rm ex}\right]&=(V_0/2)\psi_{R/L}+V_{p}\psi_{L/R},\\
\left[\psi_{R/L},H_{\rm int}\right]&=\bigr[G_1\pm E_m(|\alpha|^2-|\beta|^2)\bigr]\psi_{R/L}\\
&-2E_s\bigr[\Re(\alpha\beta^*)\pm i\Im(\alpha\beta^*)\bigr]\psi_{L/R},
\end{split}
\label{eqn:commutator2}
\end{equation}
where $V_p=\frac{V_0}{2}\cos\theta\sin\theta$ represents the coupling strength induced by external perturbation, $G_1 = \frac{n}{4}(g_s+g_a)$, $G_2 = \frac{n}{4}(g_s-g_a)$, $E_m = G_2 \cos^22\theta$, and $E_s = 2 G_1 \cos^2\theta \sin^2\theta$. From the last commutator we can see that when $\alpha\beta\neq0$, the interacting Hamiltonian also contributes to the coupling between $\psi_L$ and $\psi_R$. This is because in this case the condensate wave function is a superposition of the ground states at $k_x=\pm k_m$, which breaks translational symmetry and leads to a periodic density profile with spatial period $\delta x=\pi/k_m$ in the condensate. Accordingly, the interacting Hamiltonian $H_{\rm int}$ also breaks translational symmetry and can drive transitions between the two ground phases by transferring momentum $2k_m$ between them.

With the above results we can now write down the equation of motion in the space of $\psi_R$ and $\psi_L$:
\begin{equation} \label{EOM}
i \frac{d}{dt} \left(
\begin{array}{c}
\alpha(t) \\
\beta(t)
\end{array} \right) =
{\cal H}_{\rm eff} \left(
\begin{array}{c}
\alpha(t) \\
\beta(t)
\end{array} \right),
\end{equation}
where the effective two-state nonlinear Hamiltonian is given by
\begin{eqnarray}\label{effectiveHamiltonian}
{\cal H}_{\rm eff}&=&E^-_{\rm kin}+\frac{V_0}{2}+G_1+E_m(|\alpha|^2 - |\beta|^2)\tilde s_z\nonumber\\
&&+2E_s\bigr[\Re(\alpha\beta^*)\tilde s_x-\Im(\alpha\beta^*)\tilde s_y\bigr]+V_p\tilde s_x.
\end{eqnarray}
Here $\tilde s_{x,y,z}$ are Pauli matrices acting on the pseudospin space spanned by the two magnetized states. It can be seen that $E_m$ and $E_s$ in the above formula characterize the interaction energies for the magnetized phase and stripe phase, respectively. Actually, it is easy to verify that without external perturbation, the magnetized phase, say with $\alpha=1$ and $\beta=0$, is an eigenstate of $H_{\rm eff}$ with energy $E=E_0+E_m$, where $E_0=E^-_{\rm kin}+\frac{V_0}{2}+G_1$, and the energy of a stripe state with $|\alpha|=|\beta|=1/\sqrt{2}$ is $E=E_0+E_s$. The BEC is initialized in one of the magnetized phases if $E_m<E_s$, while in the opposite case the BEC will be initialized in a stripe phase.\\

\begin{figure*}[t]
\includegraphics[width=1.7\columnwidth]{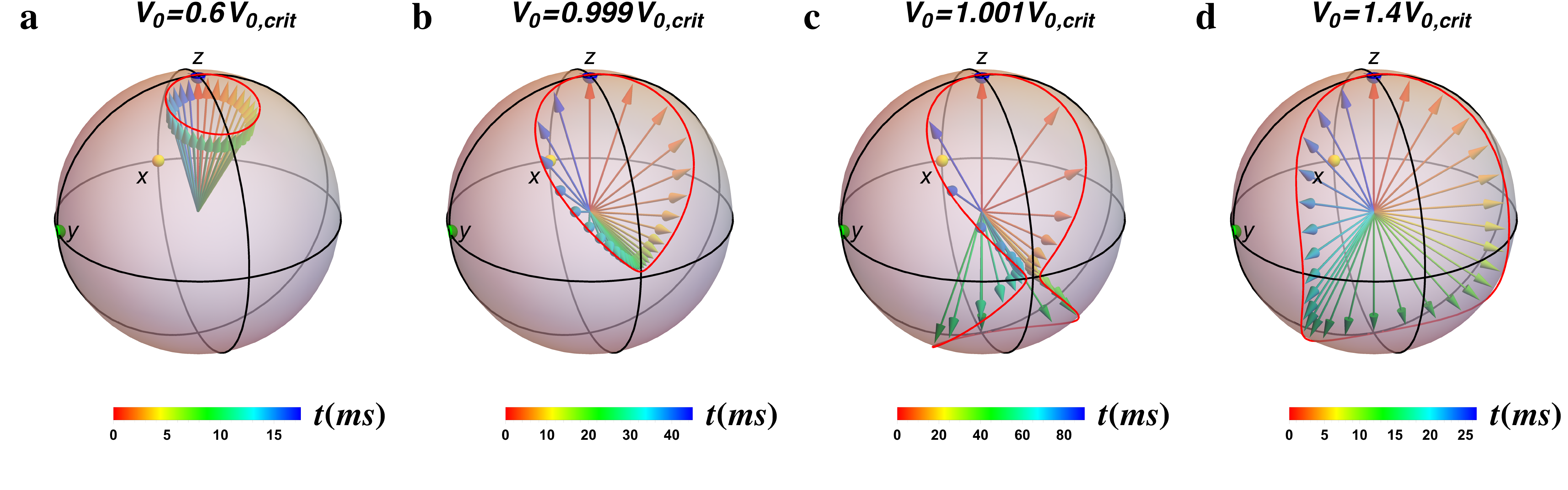}
\caption{{\bf Pseudospin spin evolution on the Bloch spherical surface.} The arrows depict the direction of pseudospin $\langle\bold{\tilde s}(t)\rangle$ during a single evolution period which starts from the north point, with the color quantifying the evolution time. {\bf a.} The external perturbation is below the critical value, with $V_0=0.6V_{0,\rm crit}$, where $V_{0,\rm crit}=4V_{p,\rm crit}/\sin2\theta$. The evolution path is nearly a circle for the small perturbation regime. {\bf b.} For the external perturbation below but close to the critical value, with $V_0=0.999V_{0,\rm crit}$, the evolution path exhibits a ``heart" shape. At the southmost point the pseudospin vector points closely to $-x$ direction, and corresponds to the maximum probability value $|\beta^2_{\rm max}|\lesssim1/2$ of the magnetized phase at $k=-k_m$. {\bf c.} The pseudospin evolution with $V_0=1.001V_{0,\rm crit}$. A full transition from the initial magnetized phase to the one at $k=-k_m$ occurs in this case. The evolution path exhibits an ``8" shape, and pass through the equatorial ($x-y$) plane through the point close to $-x$ axis. {\bf d.} The perturbation $V_0=1.4V_{0,\rm crit}$. Further increasing the perturbation broadens the evolution path. Other parameters are taken as $g_sn=1.0k_0^2$ and $\Omega=0.3k_0^2$ in the numerical simulation.}
\label{Spindynamics}
\end{figure*}

\noindent{\bf Pseudospin Evolution.}-- We numerically solve the nonlinear equation of motion with the initial condition $\alpha(0)=1$ and $\beta(0)=0$. The real spin evolution can be easily obtained from its relation to the pseudospin dynamics $\langle\sigma_z(t)\rangle=\langle\tilde s_z(t)\rangle\cos2\theta$, with $\langle\tilde s_z(t)\rangle=|\alpha|^2-|\beta|^2$, $\langle\sigma_x(t)\rangle=\sin2\theta/2$, and $\langle\tilde\sigma_y\rangle=0$. Fig.~\ref{Spindynamics} shows the pseudospin evolution, from which one can determine that time-dependent probability, given by $|\beta(t)|^2=[1-\langle\tilde s_z(t)\rangle]/2$, of the magnetized state at $k_x=-k_m$ with different perturbation amplitudes and under experimentally achievable parameter regimes. From the numerical results, we find the following novel features.

First of all, a minimum value of the perturbation $V_{p,\rm crit}$, which relates to the critical amplitude of $V_{\rm ex}$ by $V_{0,\rm crit}=4V_{p,\rm crit}/\sin2\theta$ and depends on the condensate interaction energy $g_sn$, is required to have a full transition from one magnetized phase to another. When the perturbation strength is below this critical value, the pseudospin evolves only within the upper half spherical surface and the maximum probability of the magnetized phase at $k_x=-k_m$ is less than $1/2$, i.e. $|\beta^2_{\rm max}|<1/2$ [Fig.~\ref{Spindynamics} (a,b)]. On the other hand, once the the perturbation exceeds the critical value, the pseudospin can pass through the equatorial ($x-y$) plane, where the condensate is in a stripe state, and the maximum transition ratio jumps from $|\beta^2_{\rm max}|<1/2$ to $|\beta^2_{\rm max}|=1$, thus a full transition can occur [Fig.~\ref{Spindynamics} (c,d)]. From Fig.~\ref{Spindynamics} (b,c) we can see a sudden change in the pseudospin evolution $\langle\tilde{\bold s}(t)\rangle$ from $V_p\lesssim V_{p,\rm crit}$ to $V_p\gtrsim V_{p,\rm crit}$, showing the interesting critical behavior of dynamics for the transitions between the two many-particle ground states.

It can be seen that the period of pseudospin evolution $T_R$ exhibits an increasing function of the perturbation when $V_p<V_{p,\rm crit}$ (or $V_0<V_{0,\rm crit}$), and a decreasing function of the perturbation when $V_p>V_{p,\rm crit}$. Moreover, from Fig.~\ref{Spindynamics} (b,c) one can read that the pseudospin evolves faster around $\langle\tilde s_z\rangle\sim\pm1$ than it does around $\langle\tilde s_z\rangle\sim0$.

The existence of the critical perturbation reveals a novel quantum dynamical phase transition for the many-particle states driven by the applied perturbation~\cite{dynamicaltransition1,dynamicaltransition2,dynamicaltransition3,SI}, which cannot be obtained from a single-particle two-state system. This phenomenon reflects the nontrivial interactions effects on the transitions between many-particle states, and can be qualitatively understood in the following way. When a quantum system evolves from one state to another, it must undergo the intermediate process which is a superposition of the two quantum states. For single-particle system with resonant couplings, the superposition of the two states has exactly the same energy as any of the two quantum states. Thus a resonant oscillation is guaranteed in the whole process, and a full transition can always happen as long as the perturbation is nonzero. However, for the present many-particle interacting system, the transition from one magnetized phase to the other must undergo the intermediate stripe phase which has the energy different from (greater than) that of the magnetized phases. As a result, the transition between the two ground phases is ``detuned" from the intermediate superposition state, and then a full transition is forbidden if the coupling is not strong enough.


\section*{Effective Theory for Pseudospin Dynamics}

\begin{figure}[t]
\includegraphics[width=1.0\columnwidth]{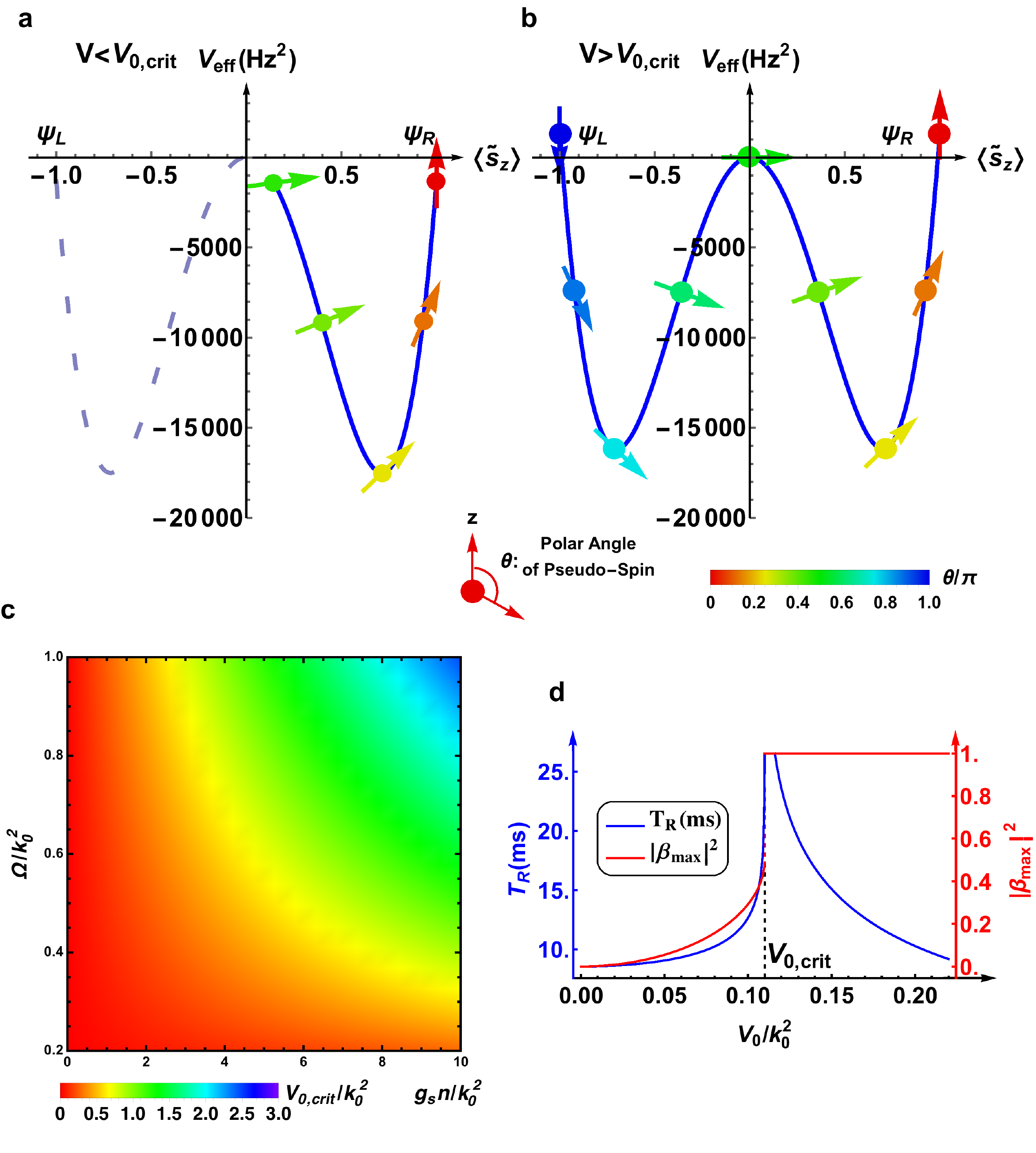}
\caption{{\bf Pseudospin spin dynamics governed by the effective double-well potential.} The initial conditions of pseudospin polarization are characterized by $\langle\tilde s_z(0)\rangle=1$ and $d\langle\tilde s_z(t)\rangle/dt|_{t=0}=0$. {\bf a.} The effective double well potential maximizes at $\langle\tilde s_z\rangle=0$ when $V_0<V_{0,\rm crit}$, and the pseudospin cannot pass through the maximum point $\langle\tilde s_z\rangle=0$. {\bf b.} The effective double well potential maximizes at $\langle\tilde s_z\rangle=\pm1$ for $V_0>V_{0,\rm crit}$. In this case the pseudospin can pass through $\langle\tilde s_z\rangle=0$ and thus a full transition occurs. {\bf c.} The critical magnitude of external perturbation as a function of Raman coupling $\Omega$ and interaction energy $g_sn$. {\bf d.} The pseudospin evolution period $T_R$ (red curve) and maximum transition ratio $|\beta_{\rm max}^2|$ (blue curve) versus Raman coupling $\Omega$, with the interaction energy taken as $g_sn=0.3k_0^2$.}
\label{Spin_Evolution}
\end{figure}

Now we turn to the analytical studies on the quantum spin dynamics. We shall develop an effective theory based on the equation of motion~\eqref{EOM}, with which we can reach a full understanding of the results obtained in the previous section. It is convenient to recast the equation of motion into a set of new formulas for the real variables introduced by: $\xi_1=|\alpha|^2, \xi_2=\Im(\alpha^*\beta),\xi_3=\Re(\alpha^*\beta)$. It follows that $d\xi_1/dt=2V_p\xi_2,d\xi_2/dt=(1-2\xi_1)(V_p+\Delta\xi_3)$, and $d\xi_3/dt=-2\Delta(1-2\xi_1)\xi_2$, with $\Delta=E_s-E_m$. Note that the initial condition is $\xi_1(0)=1$ and $\xi_2(0)=\xi_3(0)=0$, from which we further show that these equations can be organized and simplified into the following dynamical equations (see Supplementary Material for details~\cite{SI}):
\begin{equation}
  \begin{split}
\frac{d^2\langle\tilde s_z(t)\rangle}{dt^2}&+\frac{dV_{\rm eff}(\tilde s_z\rangle)}{d\langle\tilde s_z\rangle}=0, \\
\langle\tilde s_y(t)\rangle&=\frac{\partial_t\langle\tilde s_z(t)\rangle}{2V_p}, \\
\langle\tilde s_x(t)\rangle&=\frac{1}{\Delta}\bigr[\frac{\partial_t\langle\tilde s_y(t)\rangle}{2\langle\tilde s_z(t)\rangle}-V_p\bigr],\\
\end{split}
\label{eqn:EOM1}
\end{equation}
where $V_\text{eff}=(2V_p^2-\Delta^2)\langle\tilde s_z\rangle^2+(\Delta^2/2)\langle\tilde s_z\rangle^4$ is a $\phi^4$-type function of $\langle\tilde s_z\rangle$ and can be in the double-well (for $\Delta>\sqrt{2}V_p$) or single-well (for $\Delta<\sqrt{2}V_p$) form. The above equations provide a clear interpretation for the quantum spin dynamics of the present SO coupled interacting bosons. Namely, the spin evolution $\langle\tilde s_z(t)\rangle$ can be effectively treated as a classical particle which moves in the effective potential $V_\text{eff}(\langle\tilde s_z\rangle)$, with the initial position $\langle\tilde s_z(0)\rangle=1$ and initial velocity $d\langle\tilde s_z\rangle/dt|_{t=0}=0$ [see Fig.~\ref{Spin_Evolution} (a,b)]. From the formula~\eqref{eqn:EOM1}, we can exactly study the quantum transitions between the magnetized phases, and obtain the following important results.

First, the critical external perturbation can be read from the condition $V_{\rm eff}(1)=V_{\rm eff}(0)$, which follows that
\begin{eqnarray}\label{eqn:criticalfield1}
V_{p,\rm crit}=\frac{\Delta}{2}=\frac{1}{2}(E_s-E_m).
\end{eqnarray}
The numerical results of $V_{p,\rm crit}$ are shown in Fig.~\ref{Spin_Evolution} (c). When $V_p<V_{p,\rm crit}$, the effective potential $V_{\rm eff}$ is maximized at $\langle\tilde s_z\rangle=0$, i.e. $V_{\rm eff}(1)<V_{\rm eff}(0)$. Thus the potential at the middle point $\langle\tilde s_z\rangle=0$ becomes a barrier which the pseudo spin $\langle\tilde s_z\rangle$ cannot pass through [Fig.~\ref{Spin_Evolution} (a)]. This leads to $\langle\tilde s_z(t)\rangle>0$ or $|\beta^2(t)|<1/2$ at all the time. On the other hand, when the perturbation increases to be $V_p>V_{p,\rm crit}$, the effective potential is maximized at the starting point and $V_{\rm eff}(1)>V_{\rm eff}(0)$ [Fig.~\ref{Spin_Evolution} (b)]. In this case, the pseudo spin $\langle\tilde s_z\rangle$ can evolve from $\langle\tilde s_z\rangle=1$ to $\langle\tilde s_z\rangle=-1$, and a full transition is reached between the two magnetized phases. In general, the maximum transition ratio is given by:
\begin{eqnarray}\label{eqn:maximumratio}
|\beta^2_\text{max}|=\left\{ \begin{array}{ll}
        \frac{\Delta-\sqrt{\Delta^2-4V_p^2}}{2\Delta}, &\text{ when } V<V_{p,\rm crit},\\
        1, &\text{ when } V>V_{p,\rm crit}.
        \end{array} \right.
\end{eqnarray}
It is clear that $|\beta_\text{max}|^2<1/2$ when the perturbation is below the critical magnitude [Fig.~\ref{Spin_Evolution} (d)].

The effective theory given by Eqs.~\eqref{eqn:EOM1} enables a clear understanding of the underlying mechanism of the critical perturbation. The existence of critical external perturbation is due to the interaction induced energy mismatch between magnetized phases and the superposition of magnetized phases (stripe state), but not necessarily requiring that the stripe state has an energy higher than the magnetized phases. From the effective potential $V_{\rm eff}(\langle\tilde s_z\rangle)$ we can see that for the regime $E_s<E_m$, which implies that ground state of the system is a stripe phase, and assuming the system to be initialized in the state with a fixed momentum $k_x=k_m$, the full transition from the initial phase to another degenerate state at $k_x=-k_m$ still necessitates a perturbation exceeding the critical value $V_{p,\rm crit}=|E_s-E_m|/2$.

Furthermore, the period of the pseudospin oscillation $\langle\tilde s_z(t)\rangle$ can be solved exactly from the Eqs.~\eqref{eqn:EOM1}. For the two different regimes $V_p<V_{p,\rm crit}$ and $V_p>V_{p,\rm crit}$, the oscillation period takes different forms, given by $T_R=2\frac{1}{\sqrt{\Delta^2-4V_p^2}} K\left(i/r_<\right)$ and $T_R=4\frac{1}{\sqrt{4V_p^2-\Delta^2}} K\left(i/r_>\right)$, respectively. Here $K(z)$ is the complete elliptic integral of the first kind~\cite{Handbook}, with $r_<=(\Delta^2-4V_p^2)^{1/2}/(2V_p)$ and $r_>=(4V_p^2-\Delta^2)^{1/2}/\Delta$. A more intuitive understanding of the pseudospin dynamics can be acquired by expanding the oscillation period with the perturbation below and over critical value in series of $V_p$ and $1/V_p$, respectively, which gives
\begin{eqnarray}\label{eqn:period1}
T_R=\left\{ \begin{array}{ll}
        \frac{\pi}{\Delta}+ \frac{\pi V_p^2}{\Delta^3} + \frac{9\pi V_p^4}{4\Delta^5} + ..., &\text{for} \ V<V_{p,\rm crit},\\ \\
        \frac{\pi}{V_p}+\frac{\pi\Delta^2}{16V_p^3}+\frac{9\pi\Delta^4}{1024V_p^5}+..., &\text{for} \ V>V_{p,\rm crit}.
        \end{array} \right.
\end{eqnarray}
From the above results we can see that the oscillation period increases with $V_p$ when the perturbation is below critical value, while it decreases with $V_p$ if the perturbation is over critical value [c.f. numerical results in Fig.~\ref{Spin_Evolution} (d)]. This feature is clearly different from the Rabi oscillation in a single-particle two-state system, where the oscillation period decreases monotonically as perturbation strength increases. In particular, in the vicinity of $V\approx V_{p,\rm crit}$, the period can be approximated by:
\begin{eqnarray}\label{eqn:period2}
T_R\approx5.54-\frac{4}{q}\log(r)+\frac{r^2 \left[\log \left(\frac{r}{4}\right)+1\right]}{q}+O(r^2),
\end{eqnarray}
where $r=r_<$ (or $r_>$) and $q=\Delta$ (or $4V_p$) for $V_p>V_{p,\rm crit}$ (or $V_p<V_{p,\rm crit}$). The Eq.~\eqref{eqn:period2} shows that the oscillation period diverges logarithmically as the perturbation approaches to the critical value [Fig.~\ref{Spin_Evolution} (d)]. On the contrary, the oscillation become faster when $V_p$ tunes far away from $V_{p,\rm crit}$. In particular, when $V_p\ll\Delta$, the period approaches to a ``universal" value $T_R\rightarrow\pi/\Delta$ which solely depends on interaction energies of the system. On the other hand, in the limit of large perturbation $V_p\gg\Delta$, the period decreases monotonically to be $T_R\approx\pi/V_p$, implying that the spin dynamics in this regime are dominated by the external perturbation, and are more similar as those in the single-particle systems, consistent with the numerical results in the previous section.

\section*{Beyond Two-State Resonant Coupling Regime}

{\bf Multiple States Correction.}-- In the previous sections we have considered only the direct transition between two magnetized phases induced by the external perturbation $V_{\rm ex}$. On the other hand, it is easy to know that $V_{\rm ex}$ can also drive the couplings between the two ground magnetized phases and other states. The leading-order corrections include the couplings to the two lower subband states $u^{(-)}_{\pm3k_m}(\theta)$ with momenta $k_x=\pm3k_m$, and the two upper subband states $u^{(+)}_{\pm k_m}(\theta)$ with momenta $k_x=\pm k_m$ (corresponding to the Hamiltonian $H_0$). Taking into account these additional couplings yields a more complicated nonlinear dynamical equation which describes the mixing of six components and can be solved numerically (see the Supplementary Materials for details~\cite{SI}).. Note that the four additional states are detuned from the ground phases with the energy difference per particle $\Delta E\approx2k_0^2$, which is two orders greater than the critical perturbation $V_{p,\rm crit}$ in the typical parameter regime in the experiment. We expect that the density of atoms pumped to these states is very small compared with the population in the two ground states. As a result, in deriving the complete dynamical equation, one can neglect the interactions within the states $u^{(-)}_{\pm3k_m}(\theta)$ and $u^{(+)}_{\pm k_m}(\theta)$, while the interactions between them and the ground states are still considered~\cite{SI}.

\begin{figure*}[t]
\includegraphics[width=2.0\columnwidth]{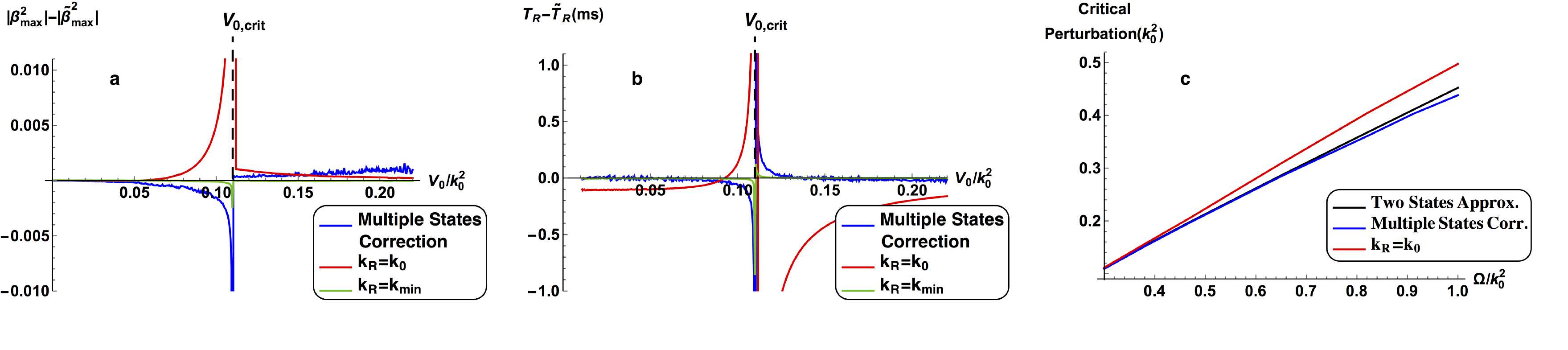}\caption{{\bf Corrections due to the couplings to multiple excited states (blue curves) and momentum mismatch (red and green curves)}. {\bf a.} Correction to the maximum transition ratio $|\beta_{\rm max}^2|-|\tilde\beta_{\rm max}^2|$. {\bf b.} Correction to the oscillation period $T_R-\tilde T_R$. {\bf c.} The corrected critical perturbation $\tilde V_{0,\rm crit}$ as a function of the Raman coupling strength. The correction for $\tilde V_{0,\rm crit}$ with $k_R=k_{\rm min}$ is nearly zero and not observable. The black curve represents the critical value $V_{0,\rm crit}$ with the two-state approximation. Other parameters are taken as $g_sn=1.0k_0^2$ and $\Omega=0.3k_0^2$ in the numerical calculation.}
\label{correction}
\end{figure*}

The numerical results of the corrections to the maximum transition ratio and oscillation period, and the corrected critical perturbation are shown with blue curves in Fig.~\ref{correction} (a-c). We find that, in all the experimentally relevant parameter regimes, the inclusion of couplings between magnetized phases and other states has negligible corrections to the spin dynamics predicted in previous sections. In particular,
when $V_0$ is tuned away from $\tilde V_{0,\rm crit}$, which is the corrected critical perturbation and is slightly smaller than $V_{0,\rm crit}$ [Fig.~\ref{correction} (c)], the maximum transition ratio $|\tilde\beta_{\rm max}|^2$ is slightly enhanced for $V_0<\tilde V_{0,\rm crit}$ compared with the result $|\beta_{\rm max}|^2$ obtained with two-state approximation [Fig.~\ref{correction} (a)]. Accordingly, the corrected oscillation period of the spin dynamics slightly increases for $V_0<\tilde V_{0,\rm crit}$ and decreases for $V_0>\tilde V_{0,\rm crit}$ [Fig.~\ref{correction} (b)]. These features suggests that the leading-order correction strengthens the effective coupling between the two magnetized phases. This phenomenon can be understood that mixing the ground phases with the additional states of high energies contributes more intermediate channels to the coupling between the two magnetized phases, and in consequence the net effective coupling ($V_0^{\rm eff}=V_0+\delta V_0$, with the correction $|\delta V_0|\ll V_0$) is enhanced. In the experimental parameter regime with $\Omega\sim0.3k_0^2$ and $g_sn\sim1.0k_0^2$~\cite{Chen2014}, the inclusion of the excited states alters the oscillation period and amplitude (for $V_0<\tilde V_{0,\rm crit}$) by less than 1\%. Therefore, we conclude that the two-state approximation employed in the previous sections well captures the quantum spin dynamics driven by the external perturbation.

{\bf Momentum Mismatch.}-- Another approximation employed in the previous studies is that we take the wave vector of standing wave light used to generate perturbation to be $k_R=k_m$. In the experiment, it might be difficult to precisely match the wave vector $k_R$ with momentum of the magnetized phases, while $k_{\rm min}$ and $k_0$ can be precisely determined. The simplest way is to put that $k_R=k_{\rm min}$ or $k_R=k_0$ in the real experiment, and then the perturbation reads $V_{\rm ex}=V_0\int d^3 \bold r \cos^2 k_Rx(\psi_\uparrow^\dagger \psi_\uparrow + \psi_\downarrow^\dagger \psi_\downarrow)$ for $t>0$. From Fig.~\ref{phasediagram} (a,b) we notice that in the typical parameter regime ($g_sn\sim1.0k_0^2$ and $0.2k_0^2<\Omega<0.6k_0^2$) for magnetized phases, the momentum mismatch $\delta k=k_R-k_m$ ($k_R=k_{\rm min}$ or $k_R=k_0$) is a small number, thus we expect that it only brings minor effects on the quantum spin dynamics. Specifically, due to the momentum mismatch, the perturbation couples the initial magnetized phase with $k_x=k_m$ to another state at $k_x=-k_m-2\delta k$, which has a small detuning from the ground magnetized state. This effect can slightly reduce the transition ratio to be $|\beta_{\rm max}^2|\lesssim1$ in the regime $V_0>\tilde V_{0,\rm crit}$. Moreover, since the spin polarization and density profile depend on the momentum of the condensate, the momentum mismatch may modify the interaction energies.

The corrections due to momentum mismatch are shown numerically with red and green curves in Fig.~\ref{correction} (a,b) with different perturbation strengths. We find that the momentum mismatch slightly affect the oscillation periods and the critical value of the perturbation. Especially, the correction is tiny when the perturbation is far away from the critical value, while it is relatively more obvious if $V_0\simeq\tilde V_{0,\rm crit}$. This is because in the latter regime, it takes relatively long time for the system to evolve through the intermediate superposition phase (stripe state). Note that in such intermediate process of passing through the stripe state, interactions play the main role in governing the quantum spin dynamics. As a result, the aforementioned modifications to the interaction energies due to the deformed spin polarization and density profile may have a more pronounced effect on the pseudospin evolution. Nevertheless, we have confirmed that in all the typical parameter regimes which are relevant for the experiment, the correction due to momentum mismatch is tiny in a large range of $V_0$.

\section*{Discussions and Conclusion}

The perturbation $V_{\rm ex}(\bold r,t)$ is a conventional spin-independent lattice potential applied at $t>0$. Differently from the spin-flip Raman coupling used to creat SO coupling, the periodic perturbation $V_{\rm ex}$ can be generated by a standing-wave light with relatively large detuning, and applying this perturbation should not induce additional heating effect. For the alkali atoms, a spin-independent optical dipole potential can be readily generated with a $\pi$-polarized light.

In experiment, the transition between the magnetized phases (pseudospin dynamics) can be detected by measuring the evolution of spin polarization of the condensate according to the relation $\langle\sigma_z(t)\rangle=\langle\tilde s_z(t)\rangle\cos2\theta$. In the typical parameter regimes with $g_sn\sim1.0k_0^2$ and $0.2k_0^2<\Omega<0.6k_0^2$, we can verify that $k_m=k_{\rm min}\simeq k_0$, and $\cos2\theta\lesssim1.0$. Thus the pseudospin polarization along $z$ axis nearly equals the spin polarization of the BEC.

The present study is not restricted by the finite temperature effect, atom loss, or the existence of a weak trapping potential. Actually, the finite temperature effect and atom loss may cause the change in the condensate density $n$, which effectively modifies the interaction energies characterized by $g_sn$ and $g_an$. As a result, these effects can quantitatively, but not qualitatively affect the critical magnitude of the perturbation. Similarly, the trapping potential in the real experimental studies may also quantitatively affect the predicted threshold of external perturbation. In the presence of a weak trapping potential along $x$ direction, the magnetized phase is not a state with fixed momentum, but has a small momentum distribution around $k_x=\pm k_m$. In this way, the effective coupling between such two phases can be generically modified compared with the case without trapping potential. Nevertheless, the main results predicted in the present work will not be changed.

In conclusion, we uncovered nontrivial quantum spin dynamics in a SO coupled Bose-Einstein condensate which has two degenerate many-particle ground states with opposite finite momenta and being resonantly coupled to each other by an external perturbation. Due to the particle-particle interactions, a novel quantum dynamical phase transition is predicted that the transition of the condensate from one ground phase to the other is forbidden when the strength of external perturbation is below a critical value, and occurs only when the perturbation exceeds such critical strength. This phenomenon is in sharp contrast to the quantum dynamics in a single-particle two-level system. We developed an effective theory and showed the exact solutions to the quantum spin dynamics, which enable a full understanding of the underlying mechanism of the predicted dynamical phase transition. It is noteworthy that the present study can be generalized to the large spin system with 1D synthetic SO coupling~\cite{syntheticdimension1,syntheticdimension2,syntheticdimension3}, in which case the $SU(N)$ or $SO(N)$ symmetric interactions of atoms may generically have nontrivial effects on the quantum spin dynamics. This work opens an important avenue in the study of novel quantum spin dynamics in the SO coupled systems with interaction, which on one hand can broaden the basic understanding of new many-body physics and, on the other hand, may have interesting applications to the quantum control and quantum information science. The high feasibility of the present study will attract experimental efforts in the near future.

\section*{Acknowledgement}

We appreciate the valuable discussions with Gediminas Juzeli\={u}nas, Youjin Deng, Shuai Chen, and Vincent W. Liu. This work
is supported in part by the One-Thousand-Young-Talent Program of China.

\noindent

\onecolumngrid

\renewcommand{\thesection}{S-\arabic{section}}
\renewcommand{\theequation}{S\arabic{equation}}
\setcounter{equation}{0}  
\renewcommand{\thefigure}{S\arabic{figure}}
\setcounter{figure}{0}  

\section*{\Large\bf Supplementary Information}

\section*{Appendix A: Derivation of simplified equation of motion}
\subsection*{Pseudo-spin}
The pseudo-spin $\langle{\bold{\tilde{s}}}(t)\rangle$ of the system is defined by (Pauli matrices acting on the pseudospin spanned by the two magnetized states):
\begin{align*}
\left(\langle\tilde{s}_x\rangle,\langle\tilde{s}_y\rangle,\langle\tilde{s}_z\rangle\right)&=
\left( \begin{matrix}
\alpha^* & \beta^*
\end{matrix} \right)
\left(
\left( \begin{matrix}
0 & 1 \\ 1 & 0
\end{matrix} \right) ,
\left( \begin{matrix}
0 & -i \\ i & 0
\end{matrix} \right) ,
\left( \begin{matrix}
1 & 0 \\ 0 & -1
\end{matrix} \right)
 \right)
\left( \begin{matrix}
\alpha \\
\beta
\end{matrix} \right) \\
&= \left( 2\Re\left(\alpha\beta^*\right), 2\Im\left(\alpha\beta^*\right), |\alpha|^2-|\beta|^2\right).
\end{align*}
The evolution $\langle\bold{\tilde{s}(t)}\rangle$ describes the dynamics of the system.

\subsection*{Commutation relations}

With the two-state approximation, the evolution of the system can be described by $\psi(t)=\alpha(t) \psi_R + \beta(t) \psi_L$, and the dynamics of the condensate can be derived according to the equation $i\frac{d\psi(t)}{dt}=[\psi(t),H_0+H_{\rm int}+V_{\rm ex}]$. The commutation relations with respect to the single-particle Hamitlonian and external perturbation are calculated by:
\begin{equation}\label{S1}
\left[\psi(t),H_0\right]=\int d^3\bold r\left[\psi(t),\psi^\dag h_0\psi\right]=E^-_{\rm kin}\psi(t),
\end{equation}
with $E^-_{\rm kin}=\left(\frac{k_m^2}{2}-\sqrt{k_m^2k_0^2+\frac{\Omega^2}{4}}\right)$, and
\begin{eqnarray}
\left[\psi(t),V_{\rm ex}\right]&=&\int d^3\bold rV_0\cos^2k_0x\left[\psi(t),\psi^\dag(\bold r) \psi(\bold r)\right]\nonumber\\
&=&\frac{V_0}{2}\left(\psi(t)+\frac{1}{2}{\left[
\begin{matrix}
\alpha(t)(\cos\theta\psi_{-k_m\uparrow}-\sin\theta\psi_{-k\downarrow})\\
        \beta(t)(\sin\theta\psi_{-k_m\uparrow}-\cos\theta\psi_{-k\downarrow})\end{matrix} \right]}\right)\nonumber\\
&\approx&\frac{V_0}{2}\left[\psi(t)+\sin\theta\cos\theta(\alpha\psi_L+\beta\psi_R)\right].
\end{eqnarray}
In Eq.~\eqref{S1} we have denoted the single-particle Hamiltonian as $h_0=\bold k^2/2-k_0k_x\sigma_z+(\Omega/2)\sigma_x$. The commutation relations with respect to the interacting Hamiltonian are given by
\begin{eqnarray}
\left[\psi_R,H_{\rm int}\right]&=&\frac{g_s}{4}\int d^3\bold r\left[\psi_R,\psi^\dag_{\uparrow}(\bold r)\psi^\dag_{\uparrow}(\bold r)\psi_{\uparrow}(\bold r)\psi_{\uparrow}(\bold r)+\psi^\dag_\downarrow(\bold r)\psi^\dag_\downarrow(\bold r)\psi_{\downarrow}(\bold r)\psi_{\downarrow}(\bold r)\right]+
\frac{g_a}{2}\int d^3\bold r\left[\psi_R,\psi^\dag_{\uparrow}(\bold r)\psi^\dag_{\downarrow}(\bold r)\psi_{\downarrow}(\bold r)\psi_{\uparrow}(\bold r)\right],\nonumber\\
&=&\frac{g_s}{2}\int d^3\bold r e^{ik_mx}\left[\cos\theta n_{\uparrow}(\bold r)\psi_{\uparrow}(\bold r)-\sin\theta n_{\downarrow}(\bold r)\psi_{\downarrow}(\bold r)\right]+\frac{g_a}{2}\int d^3\bold r e^{ik_mx}\left[\cos\theta n_{\downarrow}(\bold r)\psi_{\uparrow}(\bold r)-\sin\theta n_{\uparrow}(\bold r)\psi_{\downarrow}(\bold r)\right]\nonumber\\
&\approx&\bigr[G_1+ E_m(|\alpha|^2-|\beta|^2)\bigr]\psi_{R}-2E_s\bigr[\Re(\alpha\beta^*)+ i\Im(\alpha\beta^*)\bigr]\psi_{L},
\end{eqnarray}
and
\begin{eqnarray}
\left[\psi_L,H_{\rm int}\right]&=&\frac{g_s}{4}\int d^3\bold r\left[\psi_R,\psi^\dag_{\uparrow}(\bold r)\psi^\dag_{\uparrow}(\bold r)\psi_{\uparrow}(\bold r)\psi_{\uparrow}(\bold r)+\psi^\dag_\downarrow(\bold r)\psi^\dag_\downarrow(\bold r)\psi_{\downarrow}(\bold r)\psi_{\downarrow}(\bold r)\right]+
\frac{g_a}{2}\int d^3\bold r\left[\psi_R,\psi^\dag_{\uparrow}(\bold r)\psi^\dag_{\downarrow}(\bold r)\psi_{\downarrow}(\bold r)\psi_{\uparrow}(\bold r)\right],\nonumber\\
&=&\frac{g_s}{2}\int d^3\bold r e^{-ik_mx}\left[\sin\theta n_{\uparrow}(\bold r)\psi_{\uparrow}(\bold r)-\cos\theta n_{\downarrow}(\bold r)\psi_{\downarrow}(\bold r)\right]+\frac{g_a}{2}\int d^3\bold r e^{-ik_mx}\left[\sin\theta n_{\downarrow}(\bold r)\psi_{\uparrow}(\bold r)-\cos\theta n_{\uparrow}(\bold r)\psi_{\downarrow}(\bold r)\right]\nonumber\\
&\approx&\bigr[G_1- E_m(|\alpha|^2-|\beta|^2)\bigr]\psi_{L}-2E_s\bigr[\Re(\alpha\beta^*)- i\Im(\alpha\beta^*)\bigr]\psi_{R}.
\end{eqnarray}
where $G_1 = \frac{n}{4}(g_s+g_a)$, $G_2 = \frac{n}{4}(g_s-g_a)$, $E_m = G_2 \cos^22\theta$, and $E_s = 2 G_1 \cos^2\theta \sin^2\theta$. In the last lines of the above equations we have applied the mean-field approximation.
Then we can now write down the equation of motion in the space of $\psi_R$ and $\psi_L$:
\begin{equation} \label{EOM}
i \frac{d}{dt} \left(
\begin{array}{c}
\alpha(t) \\
\beta(t)
\end{array} \right) =
{\cal H}_{\rm eff} \left(
\begin{array}{c}
\alpha(t) \\
\beta(t)
\end{array} \right),
\end{equation}
where the effective two-state nonlinear Hamiltonian is given by
\begin{eqnarray}\label{effectiveHamiltonian}
{\cal H}_{\rm eff}=E^-_{\rm kin}+\frac{V_0}{2}+G_1+E_m(|\alpha|^2 - |\beta|^2)\tilde s_z+2E_s\bigr[\Re(\alpha\beta^*)\tilde s_x-\Im(\alpha\beta^*)\tilde s_y\bigr]+V_p\tilde s_x.
\end{eqnarray}

\subsection*{Equation of motion for spin dynamics}

Making use of the substitution $\alpha$ and $\beta$ to $\langle\bold{\tilde{s}}\rangle$, the equation of motion becomes
\begin{equation}
\frac{d\langle\tilde{s}_z\rangle}{dt}= 2\left(\frac{d\alpha^*}{dt}\alpha+c.c.\right) = -V_p \langle\tilde{s}_y\rangle \nonumber
\end{equation}
and
\begin{equation}
\begin{split}
i\frac{d\langle\tilde{s}_y\rangle}{dt}+\frac{d\langle\tilde{s}_x\rangle}{dt} &= 2\left(\frac{d\alpha}{dt}\beta^* + \frac{d\beta^*}{dt}\alpha\right) \\
 &= 2\langle\tilde{s}_z\rangle\left( -\Delta \langle\tilde{s}_y\rangle + i \left( \Delta \langle\tilde{s}_x\rangle + V_p \right) \right),
\end{split} \nonumber
\end{equation}
where $\Delta = E_s-E_m$. Therefore, the original dynamical equations can be recast into a set of real differential equations, namely,
\begin{equation}
\begin{split}
\langle\tilde{s}_z\rangle'&=-2V_p\langle\tilde{s}_y\rangle, \\
\langle\tilde{s}_y\rangle'&=2\langle\tilde{s}_z\rangle(V_p+\Delta\langle\tilde{s}_x\rangle), \\
\langle\tilde{s}_x\rangle'&=-2\Delta\langle\tilde{s}_z\rangle\langle\tilde{s}_y\rangle,
\end{split}
\end{equation}
where $\langle\tilde{s}_i\rangle'=d\langle\tilde{s}_i\rangle/dt$ ($i=x,y$ or $z$). The equation for $\langle\tilde{s}_z\rangle$ can be simplified by substituting the other two equations of $\langle\tilde{s}_{x,y}\rangle$ into it
\begin{equation}
\begin{split}
\langle\tilde{s}_z\rangle \langle\tilde{s}_z\rangle'''+4\Delta^2 \langle\tilde{s}_z\rangle^3 \langle\tilde{s}_z\rangle'- \langle\tilde{s}_z\rangle'' \langle\tilde{s}_z\rangle'&=0 \\
\left(\frac{\langle\tilde{s}_z\rangle''}{\langle\tilde{s}_z\rangle}+2\Delta^2\langle\tilde{s}_z\rangle^2\right)'&=0 \\
\left(\frac{\langle\tilde{s}_z\rangle''(0)}{\langle\tilde{s}_z\rangle(0)}+2\Delta^2\langle\tilde{s}_z\rangle(0)\right)\langle\tilde{s}_z\rangle-2\Delta^2\langle\tilde{s}_z\rangle^3&=\langle\tilde{s}_z\rangle''.
\end{split}
\end{equation}

If the system starts at $\tilde{s}_z(0)=\pm1$, which corresponds to a magnetized phase, the equation of motion can then be written as
\begin{equation}
\begin{split}
\langle\tilde{s}_z\rangle''&=\left(-4V_p^2+2\Delta^2\right)\langle\tilde{s}_z\rangle-2\Delta^2\langle\tilde{s}_z\rangle^3\\
&=-\frac{d}{d\langle\tilde{s}_z\rangle}\left(\left(2V_p^2-\Delta^2\right)\langle\tilde{s}_z\rangle^2+\frac{\Delta^2}{2}\langle\tilde{s}_z\rangle^4\right).
\end{split}\label{EOMofSz}
\end{equation}
Furthermore, the other two components of the pseudo-spin are governed by the following equations:
\begin{equation}
\begin{split}
\langle\tilde{s}_y\rangle &= -\frac{\langle\tilde{s}_z\rangle'}{2V_p} \\
\langle\tilde{s}_x\rangle &= \frac{1}{\Delta}\left(\frac{\langle\tilde{s}_y\rangle'}{2\langle\tilde{s}_z\rangle}-V_p\right).
\end{split}
\end{equation}

\section*{Appendix B: Analytic solutions to the equation of motion}

\subsection*{The full solution for pseudo-spin dynamics}
The dynamical equations of the pseudospin can be solved exactly. If starting with the magnetized phase at $k_x=k_m$, these solutions are
\begin{eqnarray}
\begin{array}{ccl}
\langle\tilde{s}_x(t)\rangle &=& -\frac{2V_p\Delta}{4V_p^2-\Delta^2}  \frac{\text{sn}^2\left(t\sqrt{4V_p^2-\Delta^2},\sqrt{\frac{\Delta^2}{\Delta^2-4V_p^2}} \right)}{\text{dn}^2\left(t\sqrt{4V_p^2-\Delta^2},\sqrt{\frac{\Delta^2}{\Delta^2-4V_p^2}} \right)}, \\
\langle\tilde{s}_y(t)\rangle &=& \frac{2V_p}{\sqrt{4V_p^2-\Delta^2}} \frac{\text{sn}\left(t\sqrt{4V_p^2-\Delta^2},\sqrt{\frac{\Delta^2}{\Delta^2-4V_p^2}} \right)}{\text{dn}^2\left(t\sqrt{4V_p^2-\Delta^2},\sqrt{\frac{\Delta^2}{\Delta^2-4V_p^2}} \right)}, \\
\langle\tilde{s}_z(t)\rangle &=& \text{cd}\left(t\sqrt{4V_p^2-\Delta^2},\sqrt{\frac{\Delta^2}{\Delta^2-4V_p^2}} \right),
\end{array}
\end{eqnarray}
where $\text{cn, dn, sd}$ are Jacobi Elliptic Functions~\cite{Handbook}.
In order to solve the period of the pseudospin oscillation, we consider two different situations. For $V_p<V_{p,\text{crit}}$, the oscillation period is determined through $\langle\tilde{s}_{z}(t)\rangle'=0$. This means that $\text{nd}(t\sqrt{4V_p^2-\Delta^2},\sqrt{\frac{\Delta^2}{\Delta^2-4V_p^2}})=0$, 
which gives $T_R=2\frac{1}{\sqrt{\Delta^2-4V_p^2}} K\left(i/r_<\right)$. Here $K(z)$ is the complete elliptic integral of the first kind~\cite{Handbook}, with $r_<=(\Delta^2-4V_p^2)^{1/2}/(2V_p)$. Accordingly, for $V_p>V_{p,\text{crit}}$, the oscillation period is determined through $\langle\tilde{s}_{z,sol}\rangle=0$, i.e. $\text{cd} (t\sqrt{4V_p^2-\Delta^2},\sqrt{\frac{\Delta^2}{\Delta^2-4V_p^2}})=0$. This gives that $T_R=4\frac{1}{\sqrt{4V_p^2-\Delta^2}} K\left(i/r_>\right)$, with $r_>=(4V_p^2-\Delta^2)^{1/2}/\Delta$. Therefore, the period of the system $T_R$ is
\begin{eqnarray}
T_R=\left\{
\begin{array}{cc}
\frac{2}{\sqrt{\Delta^2-4V_p^2}}K\left(\frac{2iV_p}{\sqrt{\Delta^2-4V_p^2}}\right) & \text{for } V_p<V_{p,\text{crit}},\\ \\
\frac{1}{\sqrt{4V_p^2-\Delta^2}}K\left(\frac{i\Delta}{\sqrt{4V_p^2-\Delta^2}}\right) & \text{for } V_p>V_{p,\text{crit}}.
\end{array}
\right.
\end{eqnarray}
The power series expansion for $K(z)$ takes the form:
\begin{eqnarray}
K(z) =
\left\{ \begin{array}{lcr}
\displaystyle\frac{\pi}{2}\sum_{n=0}^\infty \left( \frac{2n!}{2^{2n} \left(n!\right)^2 }\right)^2 z^{2n} & \text{for} & -1<z^2<1\\ \\
\displaystyle\sum_{n=0}^\infty \left( \frac{2n!}{2^{2n} \left(n!\right)^2 }\right)^2 \mathcal{K}_n(z) & \text{for} & z^2<-1,
\end{array} \right.
\end{eqnarray}
where $\mathcal{K}_n(z) = \left(\log(4|z|)-2\gamma_{2n}+2\gamma_n\right)/|z|^{2n+1}$ and $\gamma_n=\sum_{i=1}^n i^{-1}$.
Direct substitutions of the above series yield the expansions in the main text.

\subsection*{Order parameter of the dynamical Phase Transition}

We define the time average of $\langle \tilde{s}_z(t) \rangle$ over an oscillation period as the order parameter of the dynamic phase transition:
\begin{eqnarray}
\bar{M}\left(V_0\right)&=&\frac{1}{T_R}\int_0^{T_R} \langle \tilde{s}_z(t)\rangle dt\nonumber\\
&=&
\left\{ \begin{array}{ll}
\frac{\pi}{2}\frac{\sqrt{\frac{\Delta^2-4V_p^2}{\Delta^2}}}{K\left(\sqrt{\frac{-4V_p^2}{\Delta^2-4V_p^2}}\right)}, & \text{for } V_0<V_{0,\text{crit}},\\ \\
0, & \text{for } V_0>V_{0,\text{crit}}.
        \end{array} \right.
\end{eqnarray}
The critical behavior of the system can be investigated through the expansion of the order parameter $\bar{M}$ around the transition point $V_0=V_{0,\text{crit}}$. Then we find that
\[
\bar{M} =  \frac{\pi}{3\log{2}-\log{\left(1-2V_p/\Delta\right)}}\text{\ \ \ \ for } V_0 \lesssim V_{0,\text{crit}},
\]
which shows that the system undergoes a continuous transition when one tunes the perturbation strength $V_0$ through the critical point $V_{0,\text{crit}}$.
\begin{figure}[t]
\includegraphics[width=0.6\columnwidth]{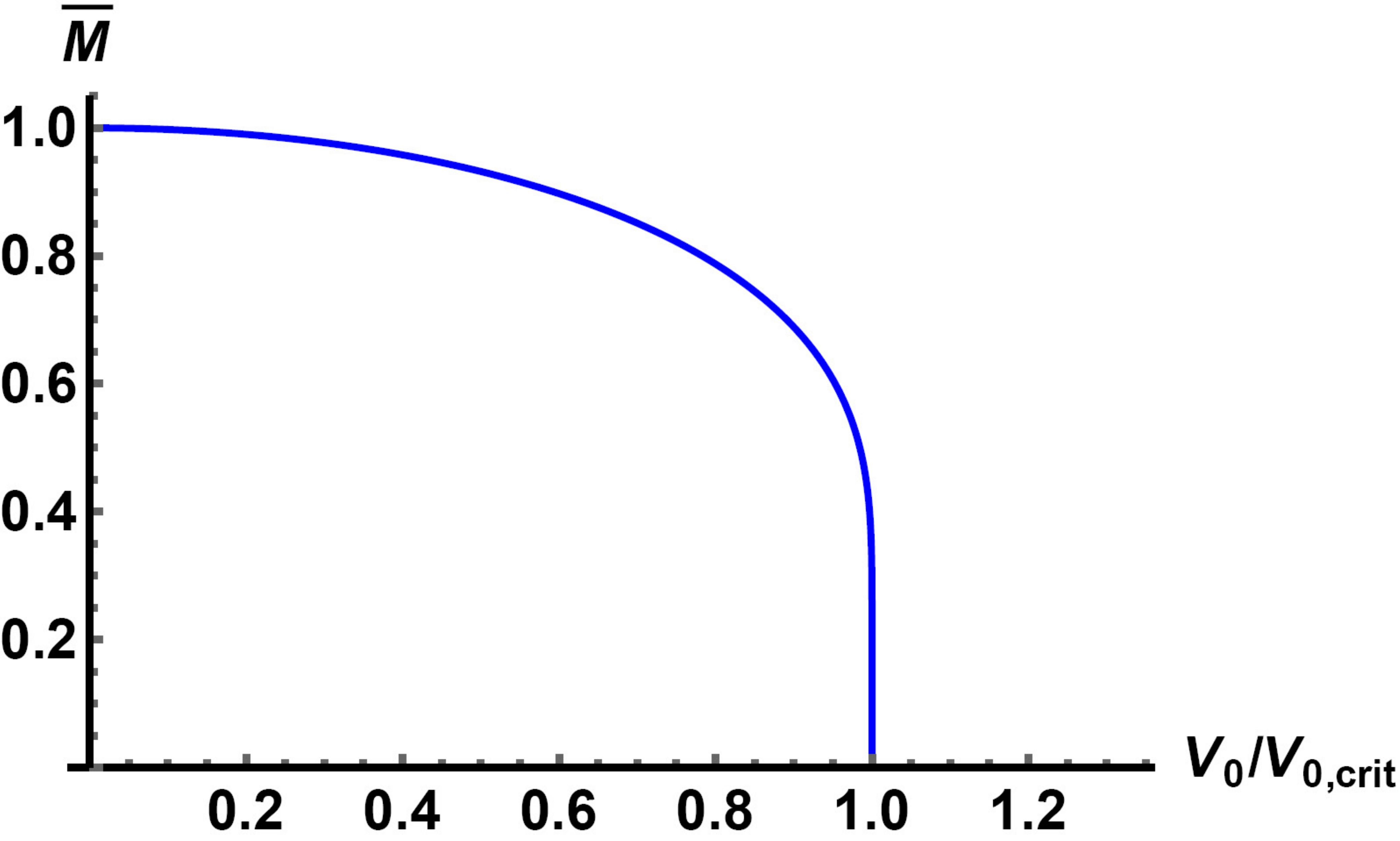}
\caption{{\bf Order parameter in dynamic phase transition.} The order parameter versus perturbation strength $V_0/V_{0,\text{crit}}$, with $V_{0,\rm crit}=\Delta/2$. The graph is independent of the choice of $n$ and $\Omega$}
\label{orderparameter}
\end{figure}

\section*{Appendix C: Beyond two-state approximation}

\subsection*{Multiple states correction}

Let $\psi_1$ and $\psi_2$ be the single-particle eigenstates with momentum $k_x=k_m$ and $k_x=-k_m$ in the upper band, respectively; $\psi_3$ ($\psi_4$) be the states with momentum $3k_m$ ($-3k_m$) in the lower band. We denote $\theta'$ as the angle describing the spin components of the states $\psi_3$ and $\psi_4$. The commutation relations are given by
\begin{equation}
\begin{split}
[\psi_{1,2},H_0]&=\left(\frac{k_m^2}{2}+\sqrt{k_m^2k_0^2+\frac{\Omega^2}{4}}\right)\psi_{1,2}; \\
[\psi_{3,4},H_0]&=\left(\frac{9k_m^2}{2}-\sqrt{9k_m^2k_0^2+\frac{\Omega^2}{4}}\right)\psi_{3,4}.
\end{split}
\end{equation}
For resonant oscillation, the second-quantized form for the  perturbation is
\begin{equation}
\begin{split}
V_\text{ex}=V_0 \int d^3k'  \frac{1}{2}\psi_{k'\sigma}^\dagger \psi_{k'\sigma}+\frac{1}{4}\left( \psi_{k'-2k_m\sigma}^\dagger \psi_{k'\sigma} + \psi_{k'+2k_m\sigma}^\dagger \psi_{k'\sigma} \right).
\end{split}
\end{equation}
By taking into account the additional states, the commutators concerning $V_\text{ex}$ are now given by
\begin{equation}
\begin{split}
[\psi_R,V_\text{ex}]&=\frac{V_0}{2}\psi_R + \frac{V_0}{4}\left[  \sin2\theta\psi_L+\cos2\theta\psi_2+
\cos\left(\theta-\theta'\right)\psi_3\right]; \\
[\psi_L,V_\text{ex}]&=\frac{V_0}{2}\psi_L + \frac{V_0}{4}\left[  \sin2\theta\psi_R-\cos2\theta\psi_1+
\cos\left(\theta-\theta'\right)\psi_4\right]; \\
[\psi_1,V_\text{ex}]&=\frac{V_0}{2}\psi_1 + \frac{V_0}{4}\left[  \sin2\theta\psi_2-\cos2\theta\psi_L+
\sin\left(\theta-\theta'\right)\psi_3\right]; \\
[\psi_2,V_\text{ex}]&=\frac{V_0}{2}\psi_2 + \frac{V_0}{4}\left[  \sin2\theta\psi_1+\cos2\theta\psi_R-
\sin\left(\theta-\theta'\right)\psi_4\right]; \\
[\psi_3,V_\text{ex}]&=\frac{V_0}{2}\psi_3 + \frac{V_0}{4}\left[  \cos\left(\theta-\theta'\right)\psi_R+
\sin\left(\theta-\theta'\right)\psi_1\right]; \\
[\psi_4,V_\text{ex}]&=\frac{V_0}{2}\psi_4 + \frac{V_0}{4}\left[  \cos\left(\theta-\theta'\right)\psi_L-
\sin\left(\theta-\theta'\right)\psi_2 \right]. \nonumber
\end{split}
\end{equation}
In the calculation, we have ignored all other states which are of higher energies than $\psi_1$ to $\psi_4$. This approximation can be justified since we shall show that the multiple states correction is indeed negligible.

The interacting term
\begin{equation}
\begin{split}
H_\text{int}&=\frac{1}{2}\int d^3r g_a \left( \psi^\dagger_\uparrow \psi^\dagger_\downarrow \psi_\downarrow \psi_\uparrow\right)+\frac{g_s}{2}\left( \psi^\dagger_\uparrow \psi^\dagger_\uparrow \psi_\uparrow \psi_\uparrow + \psi^\dagger_\downarrow \psi^\dagger_\downarrow \psi_\downarrow \psi_\downarrow \right).
\end{split}
\end{equation}
The commutation relations are
\begin{equation}
\begin{split}
[\psi_R,H_\text{int}]=
&G_1\big(\psi_R+\alpha\beta^*\sin^22\theta\psi_L+\alpha\beta^*\sin2\theta\cos2\theta\psi_2+\alpha^*\beta\sin2\theta\cos\left(\theta-\theta'\right)\psi_3\big)\\
&+G_2\big(\cos^22\theta\left(|\alpha|^2-|\beta|^2\right)\psi_R+\cos2\theta\sin2\theta\left(|\alpha|^2-|\beta|^2\right)\psi_1\big)\\
[\psi_L,H_\text{int}]=
&G_1\big(\psi_L+\alpha^*\beta\sin^22\theta\psi_R-\alpha^*\beta\sin2\theta\cos2\theta\psi_1+\alpha\beta^*\sin2\theta\cos\left(\theta-\theta'\right)\psi_4\big)\\
&+G_2\big(\cos^22\theta\left(|\beta|^2-|\alpha|^2\right)\psi_L+\cos2\theta\sin2\theta\left(|\alpha|^2-|\beta|^2\big)\psi_2\right)\\
[\psi_1,H_\text{int}]=
&G_1\big(\psi_1+\alpha\beta^*\sin^22\theta\psi_2-\alpha\beta^*\sin2\theta\cos2\theta\psi_L+\alpha^*\beta\sin2\theta\sin\left(\theta-\theta'\right)\psi_3\big)\\
&+G_2\big(\cos^22\theta\left(|\beta|^2-|\alpha|^2\right)\psi_1+\cos2\theta\sin2\theta\left(|\alpha|^2-|\beta|^2\big)\psi_R\right)\\
[\psi_2,H_\text{int}]=
&G_1\big(\psi_2+\alpha^*\beta\sin^22\theta\psi_1+\alpha^*\beta\sin2\theta\cos2\theta\psi_R-\alpha\beta^*\sin2\theta\sin\left(\theta-\theta'\right)\psi_4\big)\\
&+G_2\big(\cos^22\theta\left(|\alpha|^2-|\beta|^2\right)\psi_2+\cos2\theta\sin2\theta\left(|\alpha|^2-|\beta|^2\right)\psi_L\big)\\
[\psi_3,H_\text{int}]=
&G_1\big(\psi_3+\alpha\beta^*\sin2\theta\cos\left(\theta-\theta'\right)\psi_R+\alpha\beta^*\sin2\theta\sin\left(\theta-\theta'\right)\psi_1\big)\\
&+G_2\big(\cos2\theta\cos2\theta'\left(|\alpha|^2-|\beta|^2\right)\psi_3\big)\\
[\psi_4,H_\text{int}]=
&G_1\big(\psi_4+\alpha^*\beta\sin2\theta\cos\left(\theta-\theta'\right)\psi_L-\alpha^*\beta\sin2\theta\sin\left(\theta-\theta'\right)\psi_2\big)\\
&+G_2\big(\cos2\theta\cos2\theta'\left(|\beta|^2-|\alpha|^2\right)\psi_4\big) \nonumber
\end{split}
\end{equation}
In the above calculation we have applied the following mean-field approximation
\begin{equation}
\begin{split}
\psi^\dagger_\uparrow \psi_\uparrow &\approx \frac{N}{V}\left[ |\alpha|^2\cos^2\theta + |\beta|^2\sin^2\theta+ \cos\theta\sin\theta \left( \alpha\beta^* e^{2ik_rx} + c.c \right)\right] \\
\psi^\dagger_\downarrow \psi_\downarrow &\approx \frac{N}{V}\left[ |\alpha|^2\sin^2\theta + |\beta|^2\cos^2\theta+ \cos\theta\sin\theta \left( \alpha\beta^* e^{2ik_rx} + c.c \right)\right].
\end{split}
\end{equation}
With the above results, we can write down the effective Hamiltonian in the space spanned by the six states including two magnetized phases
\[
H^{(1)}_\text{eff}=\left(\begin{array}{cccccc}
E_0+E_m \langle\tilde{s}_z\rangle & V_p+E_s \xi & E_{m2}\langle\tilde{s}_z\rangle & V_{p2}+E_{s2} \xi & V_{p4}+ E_{s4} \xi^* & 0 \\
V_p+E_s \xi^* & E_0-E_m \langle\tilde{s}_z\rangle & -V_{p2}-E_{s2}\xi^* & E_{m2}\langle\tilde{s}_z\rangle & 0 & V_{p4} + E_{s4} \xi \\
E_{m2}\langle\tilde{s}_z\rangle & -V_{p2}-E_{s2}\xi & E_1-E_m \langle\tilde{s}_z\rangle & V_p+E_s\xi & V_{p3}+E_{s3}\xi^* & 0 \\
V_{p2}+E_{s2}\xi^* & E_{m2} \langle\tilde{s}_z\rangle & V_p+E_s\xi^* & E_1+E_m\langle\tilde{s}_z\rangle & 0 & -V_{p3}-E_{s3}\xi \\
V_{p4}+E_{s4}\xi & 0 & V_{p3}+E_{s3}\xi & 0 & E_2+E_{m3}\langle\tilde{s}_z\rangle & 0 \\
0 & V_{p4}+E_{s4}\xi^* & 0 & -V_{p3}-E_{s3}\xi^* & 0 & E_2-E_{m3}\langle\tilde{s}_z\rangle
\end{array} \right),
\]
where the matrix elements read
\begin{equation}
\begin{split}
\xi &= \alpha\beta^*, \ E_0 = \frac{k_r^2}{2}-\sqrt{k_r^2k_0^2+\frac{\Omega^2}{4}}+\frac{V_0}{2}+G_1, \\
E_1 &= \frac{k_r^2}{2}+\sqrt{k_r^2k_0^2+\frac{\Omega^2}{4}}+\frac{V_0}{2}+G_1, \ E_2= \frac{9k_r^2}{2}-\sqrt{9k_r^2k_0^2+\frac{\Omega^2}{4}}+\frac{V_0}{2}+G_1,\\
V_p &= \frac{V_0}{4}\sin2\theta, \ V_{p2}= \frac{V_0}{4}\cos2\theta, \\
V_{p3} &= \frac{V_0}{4}\sin\left(\theta-\theta' \right), \ V_{p4}= \frac{V_0}{4}\cos\left(\theta-\theta' \right), \\
E_m &= G_2 \cos^2 2\theta, \ E_{m2}= G_2 \cos 2\theta \sin 2\theta, \\
E_{m3} &= G_2 \cos 2\theta \cos 2\theta', \ E_s= \frac{G_1}{2} \sin^22\theta, \\
E_{s2} &= \frac{G_1}{2} \sin2\theta \cos2\theta, \ E_{s3} =\frac{G_1}{2} \sin2\theta \sin\left(\theta-\theta'\right), \\
E_{s4} &= \frac{G_1}{2} \sin2\theta \cos\left(\theta-\theta'\right). \nonumber
\end{split}
\end{equation}

\subsection*{Momentum mismatch}

When the wave vector $k_R$ of the laser generating the perturbation lattice potential has discrepancy from $k_m$, the initial magnetized phase with momentum $k_x=k_m$ will be coupled to the state with momentum $k_x=-k_m-\delta k$ and $\delta k=2(k_R-k_m)$. Let $\psi_a$ and $\psi_b$ be the state with momentum $k_m$ and $-k_m-\delta k$ in the lower band, and $\theta''$ be the angle describing the spin component of the state carrying momentum $-k_m-\delta k$.
The relevant commutation relations are
\begin{equation}
\begin{split}
[\psi_a,H_0] &= \left( \frac{k_r^2}{2}-\sqrt{k_r^2k_0^2+\frac{\Omega^2}{4}} \right) \psi_a ,\\
[\psi_b,H_0] &= \left( \frac{(k_r+\delta k)^2}{2}-\sqrt{(k_r+\delta k)^2k_0^2+\frac{\Omega^2}{4}} \right) \psi_b,\\
[\psi_a,V_\text{ex}] &= \frac{V_0}{2}\left(\psi_a + \frac{1}{2}\sin\left(\theta+\theta''\right)\psi_b\right),\\
[\psi_b,V_\text{ex}] &= \frac{V_0}{2}\left(\psi_b + \frac{1}{2}\sin\left(\theta+\theta''\right)\psi_a\right),\\
[\psi_a,H_\text{int}] &= G_1 \left( \psi_a + \sin^2\left(\theta + \theta'' \right) \alpha \beta^* \psi_b \right)+ G_2 \big( \left( cos^22\theta |\alpha|^2 - \cos2\theta\cos2\theta'' |\beta|^2 \right) \psi_a+ \sin^2 \left( \theta-\theta''\right) \alpha\beta^* \psi_b \big),\\
[\psi_b,H_\text{int}] &= G_1 \left( \psi_b + \sin^2\left(\theta + \theta'' \right) \alpha^* \beta \psi_a \right)+ G_2 \big( \left( cos^22\theta |\beta|^2 - \cos2\theta\cos2\theta'' |\alpha|^2 \right) \psi_b \\&+ \sin^2 \left( \theta-\theta''\right) \alpha^*\beta \psi_a \big). \nonumber
\end{split}
\end{equation}
The effective Hamiltonian governing the spin dynamics of the present system reads (in the basis of $\left[ \psi_a \hspace*{5pt} \psi_b \right]^T$)
\[
H^{(2)}_\text{eff}=\left( \begin{array}{cc}
E_0 + E_{ma}|\alpha|^2-E_{mb}|\beta|^2 & V_{pa}+E_s\alpha\beta^* \\
V_{pa}+E_s\alpha^*\beta & E_{0\delta}+E_{ma}|\beta|^2-E_{mb}|\alpha|^2
\end{array} \right),
\]
where the Hamiltonian matrix elements are
\begin{equation}
\begin{split}
E_0 &= \frac{k_r^2}{2}-\sqrt{k_r^2k_0^2+\frac{\Omega^2}{4}}+\frac{V_0}{2}+G_1 ;\\
E_{0\delta} &= \frac{(k_r+\delta k)^2}{2}-\sqrt{(k_r+\delta k)^2+\frac{\Omega^2}{4}}+\frac{V_0}{2}+G_1;\\
V_p &= \frac{V_0}{4}\sin\left(\theta+\theta''\right), \ E_{ma} = G_2 \cos^22\theta ;\\
E_{mb} &= G_2 \cos2\theta\cos2\theta'', \ E_s= \frac{G_1}{2} \sin\left(\theta+\theta''\right)+\frac{G_2}{2}\sin\theta\left(\theta-\theta''\right). \nonumber
\end{split}
\end{equation}

\end{document}